\documentclass{ws-ijmpe}
\newcommand{\am }{{\footnotesize AMPT }}
\newcommand{\amm}{{\footnotesize AMPT}}

\newcommand{\fl }{fluctuations }
\newcommand{\fll}{fluctuations.}
\newcommand{\flll}{fluctuations}

\newcommand{\corr}{correlations}

\newcommand{\et}{ $\eta$ }

\newcommand{\lngq}{\(ln<G_q>\) }
\newcommand{\lnsq}{\(lnS_q\) }
\newcommand{\lnm}{\(lnM\) }
\newcommand{\lnmm}{\(lnM\)}

\begin{document}

\markboth{Shakeel Ahmad et al}{Fluctuations of voids$\ldots$  }

\catchline{}{}{}{}{}

\title{A Study of Fluctuations of Voids in Relativistic Ion-Ion Collisions }

\author{ SHAKEEL AHMAD, M.M. KHAN, SHAISTA KHAN, A. KHATUN AND M. IRFAN}

\address{Department of Physics, Aligarh Muslim University\\
Aligarh 202002, India
\\
sa\_hep@yahoo.com}



\maketitle

\begin{history}
\received{(received date)}
\revised{(revised date)}
\end{history}

\begin{abstract}
\noindent Event-by-event \fl of hadronic patterns are investigated in terms of voids by analyzing the experimental data on 4.5, 14.5 and 60A GeV/c $^{16}$O-AgBr collisions. The findings are compared with the predictions of a multi-phase transport \am model. Dependence of voids on phase space bin width is examined in terms of two lowest moments of event-by-event \fl of voids,  \(<G_q>\) and \(S_q\). The findings reveal that scaling exponent estimated from the observed power-law behavior of the voids may be used to characterize the various properties of hadronic phase transition. The results also rule out occurrence of 2$^{nd}$ order quark-hadron phase transition at the projectile energies considered.
\end{abstract}

\section{Introduction}
\noindent One  of the main goals of studying relativistic nucleus-nucleus (AA) collisions is to investigate characteristics of strongly interacting matter under extreme conditions of initial energy density and/or temperature, where formation of quark-gluon-plasma (QGP) is envisaged to take place\cite{1,2,3}. Fluctuations in physical observables of these collisions are regarded as one of the important signals of QGP formation because of the fact that in many body systems phase transition may cause significant changes in quantum \fl of an observable from its average behavior\cite{1,3}. Furthermore, presence of large  \fl in event-by-event (ebe) analysis may indicate  existence of  distinct classes of events, one with and one without QGP formation\cite{4,5}\\
\noindent Fluctuations in physical observables are essentially of two types: the ones arising due to finite multiplicity are referred to as the statistical \fll These \fl can be filtered out by considering independent emission of particles. All other \fl are of dynamical origin and are known as 'dynamical' \flll\cite{4}. Dynamical \fl can be divided into two groups: (i) \fl which do not change on ebe basis such as two-particle correlations due to Bose Einstein statistics or due to decays of resonances and (ii) \fl exhibiting ebe variation, generally called ebe \fll  Examples are: \fl in the ratio of charged to neutral particle multiplicities due to creation of Disoriented Chiral Condensates (DCC) and occurrence of jets may contribute to the ebe \flll\cite{4}.\\
\noindent A system undergoing a second order phase transition is envisaged to exhibit large \fl and long-range \corr\cite{6}. If the symmetry of the quark system created in a heavy-ion collisions is such that the quark-hadron phase transition is of the second order then large \fl are expected in the multiplicity of hadrons not only from event-to-event but also from one region to the other in the geometrical space in which emission of hadrons takes place\cite{6}. Such local hadron density \fl would result in the formation of spatial patterns involving cluster of hadrons and regions of no hadrons between adjacent clusters\cite{6,7}. The non-hadronic regions between the clusters, termed as voids, may, therefore, provide a significant insight into the \fl associated with the critical behavior of QGP phase transition\cite{7}. Hwa and Zhang\cite{8} have proposed a method to investigate \fl of spatial patterns in terms of voids. It may be revealed that only a few attempts\cite{7,9} have been made to study  \fl of voids in hadronic and ion-ion collisions at high energies. It was, therefore,  considered worthwhile to undertake a systematic study of \fl of voids by analyzing the experimental data on AA collisions at different incident energies. This would help identify some baseline contributions to the \fl of voids.\\ 

\section{Method of Analysis}
\noindent As mentioned earlier, a method of analyzing  \fl of voids has been recently proposed by Hwa and Zhang\cite{8}. Detailed description about this method are presented in refs.7,9. However, a brief description about the method of analysis requires attention hence discussed here. It may be interesting to mention that single particle density distribution in pseudorapidity ($\eta$) space is non-flat. In order to have a flat distribution, following new cumulative variable, X($\eta$) has been introduced\cite{10,11,12}:
\begin{equation}
 X(\eta)=\frac{{\int}_{{\eta}_{min}}^{\eta}\rho(\eta) d(\eta)} {{\int}_{{\eta}_{min}}^{{\eta}_{max}}\rho(\eta) d(\eta)} 
\end{equation}
where  $\eta_{min}$ and $\eta_{max}$ denote respectively the minimum and maximum values of $\eta$-range considered, whereas $\rho(\eta)$ represents single particle rapidity density. Thus, X($\eta$) is uniformly distributed between 0 and 1. The one-dimensional X($\eta$) space is divided into M bins of equal width, $\delta$(= 1/M). The multiplicity of relativistic charged particles produced in each event will vary from bin to bin. There will be bins with a few or more hadrons and also with no hadron et al. The bins with no hadron(s) are termed as the empty bins. If two or more empty bins are adjacent, these bins are cumulatively added and treated as a single void\cite{7}. If there are one or more non-empty bins present between two empty bins, then these are considered as separate voids. The size of a void is simply the number of empty bins connected to one another. Shown in Fig.1 is a typical pattern of voids in a configuration generated for M = 16. The open squares represent empty bins, while the filled ones indicate the non-empty bins. Let \(V_k\) denote the sum of the empty bins connected then the fraction of bins that  k$^{th}$ void occupies is defined as\cite{7,9};
\begin{equation}
     x_k = V_k/M
\end{equation}
Thus, for each event there is a set \(S_e(= x_1, x_2, x_3 ....)\) of void fractions that characterizes the spatial pattern. Since the pattern fluctuates from event to event, \(S_e\) can not,therefore, be used for comparing patterns in an efficient way. For comparison, Hwa and Zhang\cite{8} have defined a moment \(g_q\) for each configuration as;
\begin{equation}
g_q = {\frac{1}{m}} {\sum_{k=1}^{m} {x_{k}}^q}
\end{equation}
\noindent where, the sum is over all voids in an event, m represents the total number of voids while q is the order of the moment. The normalized G moments of order q are defined as\cite{8}
\begin{equation}
G_q = {\frac{g_q}{g_1^q}}
\end{equation}
which depend not only on q but also on M. Thus, by definition \(G_0 = G_1 = 1\). It should be mentioned that \(G_q\) having above definition are quite different from those used in studying the erraticity of rapidity gaps\cite{8,11,13} because \(x_k\) do not satisfy the sum rule. The \(G_q\) are also different from G moments used in the study of multifractality\cite{14}. \(G_q\) given by Eq.(4) is a number for every event for a given set of values of q and M. For a fixed q and M, \(G_q\) would exhibit \fl in its values from event to event and is a measure of the void patterns.\\
\noindent The ebe \fl of \(G_q\) can be examined by plotting the probability distribution, \(P(G_q)\), for a given set of events. Using such a probability distribution, several moments may be calculated. The two lowest moments are\cite{8,9};
\begin{equation}
<G_q> = {\frac{1}{N_{ev}}} {\sum_{e=1}^{N_{ev}} {G_q^{(e)}}} 
\end{equation}
and
\begin{equation}
S_q = <G_q lnG_q>
\end{equation}
where, the superscript 'e' denotes the e$^{th}$ event and N$_{ev}$ is the number of events in a data sample. The two moments, \(<G_q>\) and \(S_q\), are expected to obey power-law behavior as; 
\begin{equation}
<G_q>  \propto M^{\gamma_q}
\end{equation}
\begin{equation}
S_q  \propto M^{\sigma_q}
\end{equation}
Such a scaling behaviour implies that voids of all sizes occur at phase transition. since the moments corresponding to different q are highly correlated, the scaling exponents \(\gamma_q\) and \(\sigma_q\) are expected to depend on q in the following fashion:
\begin{equation}
\gamma_q = c_0 + cq
\end{equation}
\begin{equation}
\sigma_q = s_0 + sq
\end{equation}
\noindent Thus, the values of c and s, which represents the slopes of slopes are concise characterization of the \fl near the critical point.\\

\section{Experimental Details}
\noindent Three samples of events produced in the interactions of 4.5, 14.5 and 60A  GeV/c $^{16}$O beams with AgBr group of nuclei in emulsion  are used in the present study. These events are taken from the emulsion experiments performed by EMU01 Collaboration\cite{15,16,17,18}. The events were searched for by along-the-track scanning method, which gives a relative minimum bias sample because of its inherent high detection efficiency\cite{15,16,19,20}. The events lying within 2-5 cm from the edge of the pellicle were considered for various measurements\cite{15,19}. The tracks of the secondary particles were identified on the basis of their ionization\cite{15,16,20}. The tracks having ionization, \(I < 1.4I_0\), where \(I_0\) is the minimum ionization produced by a singly charged relativistic particle, are known as the shower tracks. The tracks having ionization in the range, \(1.4I_0 \leq I \leq 10I_0\) are referred to as  grey tracks, whereas those having ionization \(I > 10I_0\) are termed as black tracks. The numbers of grey and black tracks produced in an event are denoted by \(n_g\) and \(n_b\) respectively. The grey and the black tracks are jointly called  as heavily ionizing tracks and their number in an interaction is denoted by \(n_h =(n_b + n_g)\).  Events with \(n_h \leq 7\) are regarded to be produced either due to interactions with H or CNO group of nuclei or due to the peripheral collisions with AgBr group of nuclei, whereas the events with \(n_h \geq 8\) are exclusively due to the collisions with AgBr group of targets\cite{16,19,21}. On the basis of these criteria, events produced in the interactions of $^{16}$O ion with AgBr targets were selected for the present analysis. The numbers of 4.5, 14.5 and 60A GeV/c $^{16}$O-AgBr collisions events turn out to be 1200, 370 and 422 respectively.  It should be emphasized that the conventional emulsion technique has two main advantages over the other detectors: (i) its \(4\pi\) solid angle coverage and (ii) emulsion data are free from biases due to full phase space coverage. However, in the case of other detectors, only a fraction of charged particles are recorded due to the limited acceptance cone. This not only reduces the charged particle multiplicity but also distorts some of the event characteristics, like particle density \flll\cite{19,22}. In order to compare the results of the present study with those predicted by a multi phase transport (\amm) model, matching numbers of events were simulated using the code, ampt-v1.21-v2.21\cite{23}. The events are simulated by taking into account the percentage of interactions which occur in collisions of projectile with various targets in emulsion\cite{18}. While generating the \am event samples, the values of impact parameter are so set that the mean multiplicities of the relativistic charged particles, \(<n_s>\) turns out to be nearly equal to those obtained for the experimental data.

\section{Results and Discussion}
                 
Values of pseudorapidities of the relativistic charged particles produced in each event lying in the range, \(\Delta \eta(=\eta_c \pm 3.0)\), where \(\eta_c\) is the center of symmetry of the \et distribution, are transformed into  \(X(\eta)\) using Eq.(1). The one dimensional \(X(\eta)\) space is divided into M bins of equal width, \(\delta(= 1/M)\). The presence of voids are searched for following the approach  discussed above and the sizes of the voids are determined. Void fractions, \(x_k\), \(g_q\) moments and normalized \(G_q\) moments are calculated using Eqs.2-4. The values of \(G_q\) for q = 2-5 are calculated by varying the number of bins, M, from 16 to 96 in steps of 8. Distributions of \(G_q\) moments for q = 2 and M = 16 \& 64 for the experimental and \am data at the three incident energies are plotted in Figs.2,3. It may be noticed from these figures that \(G_q\) fluctuate from event to event and the \fl are rather more pronounced for higher number of phase space bins. It may be of interest to mention that the shapes of \(G_q\) distributions are satisfactorily reproduced by those obtained using \am event samples. Furthermore, it is evident from Figs.2 and 3 that mean value of \(G_q\) and the dispersion of \(G_q\) distribution vary with M, for a fixed q, and also with q, for a fixed M (not shown). This suggests that in order to look at simple regularities in the nature of \fl of \(G_q\), M dependence of \(<G_q>\) over all configurations must be looked into\cite{8}. \\
\noindent Variations of \lngq with \lnm for various data sets are shown in Figs.4 and 5. It is seen from the figures that \lngq increases with \lnm. The lines in the figures are due to the best fits to the data of the form:
\begin{equation}
ln<G_q> = a_1 + b_1M + c_1M^2
\end{equation}
\noindent Values of constants \(a_1\), \(b_1\), \(c_1\), appearing in the above expression are listed in Table~1. Our results reveal that variations of \lngq with \lnm are quadratic in nature for both real and \am data sets at the three incident energies. These results are somewhat different from those reported earlier\cite{7} for 200A GeV/c $^{32}$S-AgBr and 350 GeV/c $\pi^-$-AgBr collisions. These workers have observed a linear increase of \lngq with \lnmm. Hwa and Zhang\cite{8} have also suggested that \(G_q\) vs M plots on log-log scale should be linear. Data points for to M = 32 and above, displayed in Figs. 4 and 5, indicate also linear increase of \lngq with \lnm and suggest a power-law behavior expressible by Eq.(7). The values of slopes \(\gamma_q\) obtained from the best fits to the data (for M $\geq$ 32) are presented in Table~2. It may be noted from the Table that the values of \(\gamma_q\) for the experimental data are slightly higher than those for the \am events. Such  scaling behavior may be interpreted as indication of occurrence of voids of all sizes\cite{8}.\\
\noindent Since the \(G_q\) moments of various orders are highly correlated, the scaling exponent \(\gamma_q\), appearing in Eq.(7) is expected to depend on q in some simple ways. In order to examine this, variations of \(\gamma_q\) with q for the  various data sets are plotted in Fig.6. The straight lines in the figure represent the best fits to the data of the form, given by Eq.(9). The values of coefficient "c" occurring in Eq.(9) are listed in Table~3. It has been suggested\cite{8} that Eq.(9) may be regarded as a convenient parameterization of \(\gamma_q\) that allows to regard "c" as a numeric description of the scaling behavior of voids.\\
\noindent Distributions of \(G_q\), shown in Figs.2 and 3 indicate that by studying the behavior of \(<G_q>\) only a limited  information about the distribution may be extracted. However, the shape of \(G_q\) distribution would characterize the nature of ebe \fl in the distribution. A moment which quantifies the degree of these \fl is expressed as;
\begin{equation}
C_{p,q} = {\frac{1}{N}} {\sum_{e=1}^{N} {(G_q^{(e)})^p}} = \int dG_qG^p_qP(G_q)
\end{equation}
As \(C_{1,q} (=<G_q>)\) provides no information about the extent of \fl but the derivative of \(c_{p,q}\) at \(p=1\),
\begin{equation}
S_q = \frac {d}{dp} C_{p,q}{\Big|}_{p=1 } = <G_qlnG_q>
\end{equation}
is envisaged to yield maximum information regarding the ebe \fl\\
\noindent Variations of \lnsq with \lnm for the real data at the three incident energies are displayed in Fig.7, while Fig.8 exhibits similar plots for the \am events. The lines in the figures are due to the best fits to the data of the following form:
\begin{equation}
lnS_q = a_2 + b_2M + c_2M^2
\end{equation}
The values of the constants \(a_2\), \(b_2\), \(c_2\), estimated for various event samples, are presented in Table~4. These results indicate that the trend of increase of \lnsq with \lnm may be nicely reproduced by a 2$^{nd}$ order polynomial. However, like \(G_q\),  linear fits to the data points for \(M \geq 32\), shown in Figs.7 and 8 are obtained to study the power-law behavior expressed by Eq.(8). The values of slopes \(\sigma_q\) are presented in Table~5. It is observed that the values of \(\sigma_q\) for the experimental data are somewhat higher than those obtained for the \am event samples. Dependence of \(\sigma_q\) on q is displayed in Fig.9.  The straight lines in the figure are due to parameterization of Eq.(10). The values of the coefficient "s" are presented in Table~3. It may be noted from the table that the values of "s" for the real data are slightly higher than those for the \am events. It may also be noted from the table that the values of "c" and "s"  increase with increasing beam energy. It should be mentioned that the values of "c" for the experimental data lie between 0.26 and 0.46, whereas the values of "s" lie in the range 0.23 and 0.38. The values of "c" and "s" in the case of 200A GeV/c $^{32}$S-AgBr collisions have been reported\cite{7} to be \(0.56\pm0.01\) and \(0.43\pm0.02\) respectively. As a quantitative signature of 2$^{nd}$ order quark-hadron phase transition, the values of "c" and "s" are predicted\cite{8} to lie in the ranges 0.75-0.96 and 0.7-0.9 respectively. The values of the two parameters obtained in the present study are much smaller than those predicted for the 2$^{nd}$ order quark-hadron phase transition, suggesting that no such phase transition occurs at the energies considered in the present study.\\
\noindent The method of voids used in the present study is one of the several approaches for studying ebe \flll, e.g., normalized factorial moments\cite{24}, multifractals\cite{12,14}, erraticity\cite{11,25}, k-order rapidity spacing \cite{22} and transverse momentum spectra\cite{26}. Furthermore, \fl in the conserved quantities like baryon number, strangeness, electric charge, etc., have emerged as new tools to estimate degree of equilibration and criticality of the measured systems\cite{27,28}. Higher moments like variance, $\sigma^{2}$, skewness, S, kurtosis, k, etc, of conserved quantities such as net baryon, net charge and net strangeness multiplicity distributions are regarded as important tools to search for the QCD critical point at RHIC and LHC energies\cite{29,30,31,32}. Dynamical development of cooling and hadronization of QGP are studied in a simple model envisaging critical \fl in the QGP to hadronic matter and a $1^{st}$ order transition in a small finite system\cite{32}. Variations of skewness and kurtosis as functions of system's temperature and volume abundance of hadronic matter are predicted which can be investigated via a beam energy scan program using the data at RHIC energies. Non-statistical \fl of higher order moment singularities of net proton event distributions in Au-Au collisions in the energy range: \(\sqrt {s_{NN}}\) =11.5-200 GeV have been studied in the framework of parton and hadron cascade model, PACIAE\cite{29,30,31}. It has been observed\cite{30,31} that the energy dependence of higher order moments of the net proton multiplicity distributions observed using STAR data are in fine agreement with the predictions of the model\cite{30}. PACIAE model predictions relating to energy dependence of various moments of net proton multiplicity distributions are also observed to be in fine agreement with the experimental results obtained using pp data at RHIC and LHC energies\cite{29}.\\
\noindent A method to determine critical temperature, $T_{c}$, by analysing the RHIC data has been proposed\cite{30,33}. By studying the variations of $\sigma^{2}$, S and k with beam energy and comparing the findings with the lattice calculation results, the value of $T_{c}$ has been reported to be {$175_{-7}^{+1}$} MeV. However, no evidence of singularity in the study of energy dependence(for a fixed centrality) and/or centrality dependence( at a given energy) has been reported\cite{31}. In order to explore the QCD phase diagram, several  attempts\cite{34,35,36,37,38,39,40} have also been made by analyzing the experimental data on pp and AA collisions in the center-of-mass energy range,  \(\sqrt {s_{NN}} \sim\) 7.7-200 GeV; this will also help study the various interesting features of higher order moments of net proton multiplicity distributions. Although these investigations do not provide unambiguous information that S and k and the products $S\sigma$ and $k\sigma^{2}$ are sensitive to the quark-gluon-hadron phase transition or any potential QCD critical point\cite{36,37}, yet the findings help understand non QCD critical point physics effects in the observables such as effects of conservation(of electric charge, baryon number and strangeness number), finite size, decays of resonances and hadronic scattering. Moreover, from the results based on the observables a baseline may be disentangled for the observables in order to search for the critical point\cite{38}.\\
\noindent As already stated we have analyzed the experimental data collected using emulsion technique, where the charges can not be identified, study of net proton multiplicity distribution is not possible. We have, therefore, used the method of voids to study the \fl on ebe basis. It may be remarked that this method is fairly suitable for such studies using the data at relatively lower energies, AGS and SPS, because multiplicities at these energies are generally low and hence probability of occurrence of voids in the available range of rapidity will be quite large. If the multiplicities of hadrons vary significantly from one region to the other in the geometrical phase space then it becomes important to capture \fl using a detector capable of covering full phase space. Emulsion technique  is the most suitable and advantageous because it allows full phase space coverage. Thus, the results obtained in the present study may help construct a baseline to the \fl arising from non-QGP critical point physics.\\

\section{Conclusions}
\noindent Results of the present study encourage to draw the following important conclusions:
\begin{enumerate}
\item \(G_q\) distributions for different number of phase space bin, M, reveal that \(G_q\) fluctuate from event to event and the \fl are more pronounced in the case of higher values of M.\\
\item Trends of variations of \(<G_q>\) and \(S_q\) with number of bins, M, (on log-log scales) are nicely reproduced by a 2$^{nd}$ order polynomial. However, for higher values of M, these variations tend become linear. Such a linear dependence would indicate  the presence of power-law behavior, which would indicate existence of voids of all sizes.\\
\item Scaling exponents, \(\gamma_q\) and \(\sigma_q\) estimated from the power-law dependence of \(G_q\) and \(S_q\) on the order number q are observed to grow linearly. However, the slopes obtained from the plots of \(\gamma_q\) and \(\sigma_q\) as a function of plots are found to be relatively much smaller than those predicted envisaging quark-hadron phase transition.\\
\item Various characteristics of \fl of voids for the experimental data are observed to be in fine agreement with those predicted by \am model.
\end{enumerate}


\newpage
\noindent Table~1: Values of  \(a_1\), \(b_1\), \(c_1\) appearing in Eq. (10).
\begin{footnotesize}
\begin{center}

\begin{tabular}{c c c c c c } \hline
 &  Energy & q & a$_1$ & b$_1$ & c$_1$ \\  
 &  \small{(A GeV/c)} &  &  &  & \\ \hline 

 &  &  2  & -1.899 $\pm$ 0.073   & 1.039 $\pm$ 0.041  & -0.108 $\pm$ 0.005   \\
  & 4.5 &  3  & -3.741 $\pm$ 0.120  & 2.109 $\pm$ 0.070  & -0.209 $\pm$ 0.010   \\
    &  &  4  & -5.403 $\pm$ 0.166 & 3.144 $\pm$ 0.102   & -0.299 $\pm$ 0.015   \\
    &  &  5  & -7.201 $\pm$ 0.228   & 4.298 $\pm$ 0.142   & -0.403 $\pm$ 0.021   \\ \cline{2-6}
  &  &  2  & -3.357 $\pm$ 0.210   & 1.675 $\pm$ 0.112   &  -0.179 $\pm$ 0.014  \\
    & 14.5  &  3  & -4.885 $\pm$ 0.315   & 2.486 $\pm$ 0.181   & -0.237 $\pm$ 0.025 \\
    &  &  4  & -6.329 $\pm$ 0.425    & 3.305 $\pm$ 0.260   &  -0.286 $\pm$ 0.038  \\
    Expt. & &  5  & -7.997 $\pm$ 0.569   & 4.292 $\pm$ 0.362   & -0.352 $\pm$ 0.055  \\\cline{2-6}
     & & 2 &  -6.249 $\pm$ 0.310 & 2.893 $\pm$ 0.158   & -0.313 $\pm$ 0.020       \\
   &  60 &  3  & -6.662 $\pm$ 0.444   & 3.0 $\pm$ 0.246   &  -0.278 $\pm$ 0.033   \\
    &  &  4  & -7.235 $\pm$ 0.595   & 3.234 $\pm$ 0.349    &  -0.242 $\pm$ 0.049  \\
    &  &  5  & -8.083 $\pm$ 0.768   &  3.643 $\pm$ 0.468  &  -0.221 $\pm$ 0.068  \\ \cline{1-6}
 
     &  &  2  & -1.882 $\pm$ 0.061   & 1.027 $\pm$ 0.034  & -0.109 $\pm$ 0.004   \\
  & 4.5 &  3  & -4.291 $\pm$ 0.112  & 2.4 $\pm$ 0.065  & -0.253 $\pm$ 0.009   \\
   &  &  4  & -6.395 $\pm$ 0.149   & 3.662 $\pm$ 0.092   & -0.375 $\pm$ 0.013   \\ 
  &  &  5  & -8.73 $\pm$ 0.196   & 5.095 $\pm$ 0.127   &  -0.517 $\pm$ 0.019  \\\cline{2-6}
       &  &  2  & -3.781 $\pm$ 0.219   & 1.851 $\pm$ 0.114  & -0.201 $\pm$ 0.014   \\
    & 14.5  &  3  & -5.935 $\pm$ 0.358   & 2.945 $\pm$ 0.197   & -0.297 $\pm$ 0.026 \\
    &  &  4  & -7.232 $\pm$ 0.511    & 3.637 $\pm$ 0.295   &  -0.334 $\pm$ 0.041  \\
    \am & &  5  & -8.293 $\pm$ 0.703   & 4.262 $\pm$ 0.421   & -0.357 $\pm$ 0.060  \\ \cline{2-6}
      & &   2  & -5.543 $\pm$ 0.287   & 2.529 $\pm$ 0.147   & -0.269 $\pm$ 0.018   \\
   &  60 &  3  &  -5.727 $\pm$ 0.404  & 2.574 $\pm$ 0.224   & -0.233 $\pm$ 0.030   \\
    &  &  4  & -6.326 $\pm$ 0.551   & 2.893 $\pm$ 0.324   & -0.224 $\pm$ 0.046   \\
    &  &  5  &  -7.304 $\pm$ 0.741  & 3.477 $\pm$ 0.455   &  -0.245 $\pm$ 0.067   \\ \cline{1-6}
    
\end{tabular}

\end{center}
\end{footnotesize}

\newpage
\noindent Table~2: Values of slopes \(\gamma_q\) appearing in Eq. (6) for various data sets at different energies.
\begin{footnotesize}
\begin{center}
\begin{tabular}{c c c c c } \hline
 & & \multicolumn{3}{c} {Energy (A GeV/c)}\\\cline{3-5}
Type of events &  q & 4.5 & 14.5 & 60  \\  

     &  2 & 0.180 $\pm$ 0.006 & 0.250 $\pm$ 0.013 & 0.317 $\pm$ 0.015 \\
Expt.&  3 & 0.420 $\pm$ 0.016 & 0.598 $\pm$ 0.037 & 0.754 $\pm$ 0.040 \\
     &  4 & 0.689 $\pm$ 0.028 & 1.006 $\pm$ 0.071 & 1.238 $\pm$ 0.081 \\
     &  5 & 0.980 $\pm$ 0.042 & 1.455 $\pm$ 0.113 & 1.734 $\pm$ 0.136 \\[1mm] \hline

     &  2 & 0.160 $\pm$ 0.005 & 0.232 $\pm$ 0.010 & 0.307 $\pm$ 0.016 \\
\am   &  3 & 0.368 $\pm$ 0.014 & 0.569 $\pm$ 0.030 & 0.666 $\pm$ 0.038 \\
     &  4 & 0.599 $\pm$ 0.027 & 0.947 $\pm$ 0.062 & 1.095 $\pm$ 0.077 \\
     &  5 & 0.847 $\pm$ 0.043 & 1.249 $\pm$ 0.106 & 1.486 $\pm$ 0.132 \\[1mm] \hline
\end{tabular}
\end{center}
\end{footnotesize}  

\newpage 

\noindent Table~3: Values of coefficients c and s estimated using Eqs. (8) and (9) respectively.
\begin{footnotesize}
\begin{center}
\begin{tabular}{c c c c c} \hline
  
Energy   & \multicolumn{2}{c}{c}               & \multicolumn{2}{c}{s} \\ [2mm] \cline{2-5} 
(A GeV/c) &        Expt.    &   AMPT              & Expt.            & AMPT    \\ [2mm] \hline
4.5  &  0.26$\pm$0.01  & 0.22$\pm$0.01       & 0.23$\pm$0.02     & 0.18$\pm$0.02 \\ [2mm]
14.5 & 0.38$\pm$0.02  & 0.34$\pm$0.02       & 0.32$\pm$0.04     & 0.26$\pm$0.04 \\ [2mm]
60 &   0.46$\pm$0.03  & 0.38$\pm$0.03       & 0.38$\pm$0.05     & 0.29$\pm$0.05 \\ \hline 
\end{tabular}

\end{center}
\end{footnotesize}
\newpage
\noindent Table~4: Values of  \(a_2\), \(b_2\), \(c_2\) appearing in Eq.(10).
\begin{footnotesize}
\begin{table}[htbp]
\begin{center}
\begin{tabular}{c c c c c c} \hline

  & &  &  &  &  \\ 
 \raisebox{1.5ex}[0cm][0cm] & Energy & q & a$_2$ & b$_2$ & c$_2$ \\ 
 &  \small{(A GeV/c)} & & & & \\ \hline
 
&   &  2  & -10.979 $\pm$ 0.295   & 4.910 $\pm$ 0.015  & -0.545 $\pm$ 0.021   \\
  & 4.5  &  3  & -11.580 $\pm$ 0.355  & 5.920 $\pm$ 0.920  & -0.648 $\pm$ 0.026   \\
    &  &  4  & -12.346 $\pm$ 0.415 &  6.912 $\pm$ 0.235   & -0.741 $\pm$ 0.032   \\
    &  &  5  & -13.265 $\pm$ 0.475   & 7.871 $\pm$ 0.275   & -0.825 $\pm$ 0.038   \\ \cline{2-6}
  &  &  2  & -12.574 $\pm$ 0.741   & 5.361 $\pm$ 0.396   &  -0.573 $\pm$ 0.052  \\
    & 14.5   &  3  & -13.333 $\pm$ 0.931   & 6.510 $\pm$ 0.514   & -0.689 $\pm$ 0.070 \\
  Expt.  &  &  4  & -14.367 $\pm$ 1.128    & 7.569 $\pm$ 0.640   &  -0.781 $\pm$ 0.089  \\
     & &  5  & -15.382 $\pm$ 1.324   & 8.509 $\pm$ 0.766   & -0.850 $\pm$ 0.108  \\ \cline{2-6}
     & & 2 &  -13.832 $\pm$ 0.741 & 5.465 $\pm$ 0.396   & -0.536 $\pm$ 0.052       \\
   &  60 &  3  & -14.276 $\pm$ 1.168   & 6.306 $\pm$ 0.641   &  -0.616 $\pm$ 0.008   \\
    &  &  4  & -15.014 $\pm$ 1.393   & 7.137 $\pm$ 0.787    &  -0.612 $\pm$ 0.108  \\
    &  &  5  & -15.692 $\pm$ 1.601   &  7.817 $\pm$ 0.924 &  -0.695 $\pm$ 0.129  \\ \cline{1-6}
 
      &  &  2  & -10.303 $\pm$ 0.351   & 4.910 $\pm$ 0.185  & -0.493 $\pm$ 0.029   \\
  & 4.5  &  3  & -11.589 $\pm$ 0.502  & 5.900 $\pm$ 0.270  & -0.650 $\pm$ 0.035   \\
   &  &  4  & -13.751 $\pm$ 0.696   & 7.611 $\pm$ 0.386   & -0.841 $\pm$ 0.051   \\
  &  &  5  & -16.440 $\pm$ 0.928   & 9.529 $\pm$ 0.512   &  -1.052 $\pm$ 0.069  \\\cline{2-6}
       &  &  2  & -13.423 $\pm$ 0.794  & 5.550 $\pm$ 0.414  & -0.581 $\pm$ 0.053   \\
    & 14.5   &  3  & -13.271 $\pm$ 1.018   & 6.172 $\pm$ 0.550   & -0.634 $\pm$ 0.073 \\
 AMPT   &  &  4  & -13.385 $\pm$ 1.252    & 6.738 $\pm$ 0.698   &  -0.669 $\pm$ 0.095  \\
     & &  5  & -13.594 $\pm$ 1.477   & 7.27 $\pm$ 0.847   & -0.691 $\pm$ 0.118  \\ \cline{2-6}
      & &   2  & -12.14 $\pm$ 0.859   & 4.606 $\pm$ 0.459   & -0.448 $\pm$ 0.519   \\
   &  60 &  3  &  -12.063 $\pm$ 1.054  & 5.270 $\pm$ 0.583   & -0.507 $\pm$ 0.079   \\
    &  &  4  & -12.525 $\pm$ 1.283   & 6.022 $\pm$ 0.731   & -0.567 $\pm$ 0.101   \\
    &  &  5  &  -13.040 $\pm$ 1.516  & 6.72$\pm$ 0.885   &  -0.611 $\pm$ 0.126   \\ \hline
    
\end{tabular}
\end{center}
\end{table}
\end{footnotesize}


\newpage
Table~5: Values of slope \(\sigma_q\) for the experimental and \am events at different energies.
\begin{footnotesize}

\begin{center}
\begin{tabular}{c c c c c} \hline
 & & \multicolumn{3}{c} {Energy (A GeV/c)}\\\cline{3-5}
Type of events  &  q & 4.5 & 14.5 & 60  \\  \hline

     &  2 & 0.515 $\pm$ 0.021 & 0.744 $\pm$ 0.050 & 0.937 $\pm$ 0.063 \\
Expt.&  3 & 0.707 $\pm$ 0.032 & 1.023 $\pm$ 0.079 & 1.314 $\pm$ 0.098 \\
     &  4 & 0.950 $\pm$ 0.044 & 1.392 $\pm$ 0.116 & 1.682 $\pm$ 0.147 \\
     &  5 & 1.228 $\pm$ 0.059 & 1.730 $\pm$ 0.158 & 2.103 $\pm$ 0.203 \\[1mm] \hline

     &  2 & 0.488 $\pm$ 0.018 & 0.795 $\pm$ 0.045 & 0.808 $\pm$ 0.062 \\
\am   &  3 & 0.637 $\pm$ 0.029 & 1.025 $\pm$ 0.076 & 1.158 $\pm$ 0.098 \\
     &  4 & 0.833 $\pm$ 0.044 &  1.331 $\pm$ 0.118 & 1.429 $\pm$ 0.148 \\
     &  5 & 1.061 $\pm$ 0.062 & 1.588 $\pm$ 0.170 & 1.754 $\pm$ 0.207 \\[1mm] \hline
\end{tabular}

\end{center}
\end{footnotesize}

\newpage
\begin{figure}[th]
\centerline{\psfig{file=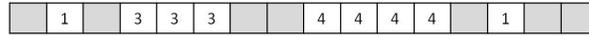,width=8cm}}
\vspace*{8pt}
\caption{Schematic representation of a typical void pattern in one dimensional space.}
\end{figure}
%
\newpage
\begin{figure}[th]
\centerline{\psfig{file=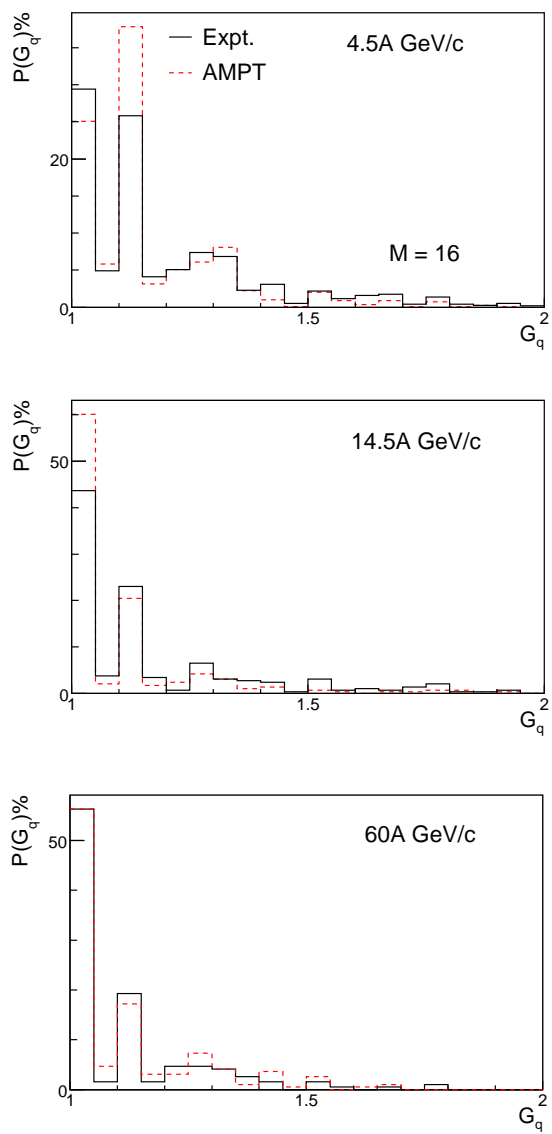,width=8cm}}
\vspace*{8pt}
\caption{\(G_q\) distributions  for q =2 and M = 16 for the real and \am data on 4.5, 14.5 and 60A GEV/c  $^{32}$S-AgBr interactions.}
\end{figure}
%
\newpage
\begin{figure}[th]
\centerline{\psfig{file=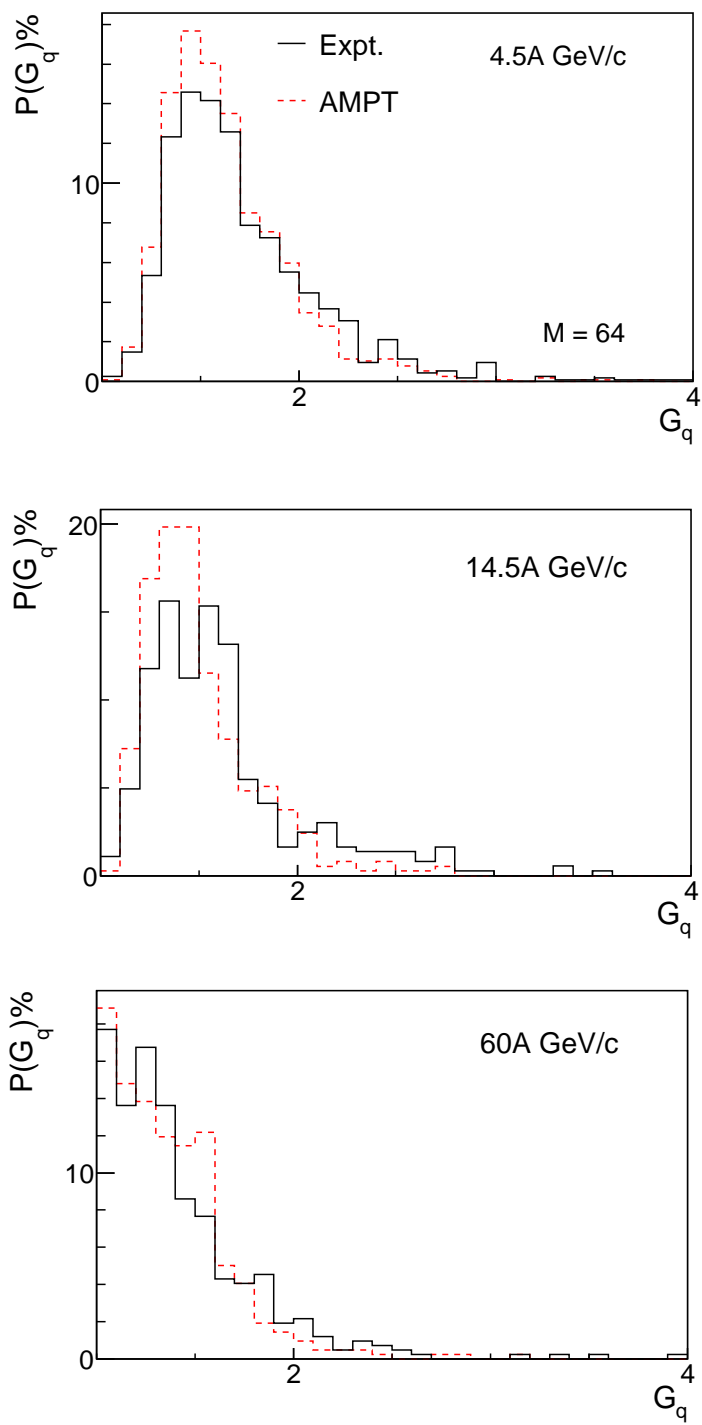,width=10cm}}
\vspace*{8pt}
\caption{ \(G_q\) distributions for q = 2 and M = 64 for the experimental and \am  data on 4.5, 14.5 and 60A GEV/c  $^{32}$S-AgBr collisions.}
\end{figure}
%
%
\newpage
\begin{figure}[th]
\centerline{\psfig{file=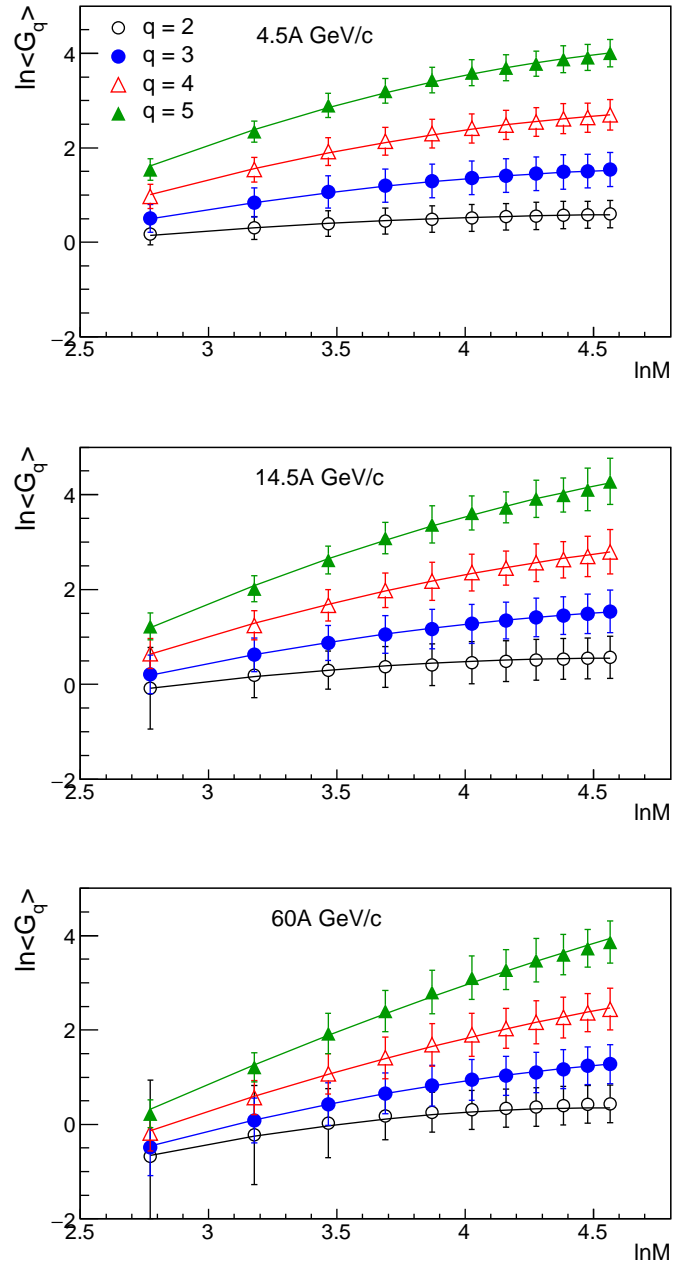,width=10cm}}
\vspace*{8pt}
\caption{Dependence of \(ln<G_q>\) on \(lnM\) for the experimental events at the three incident energies.}
\end{figure}
\newpage
\begin{figure}[th]
\centerline{\psfig{file=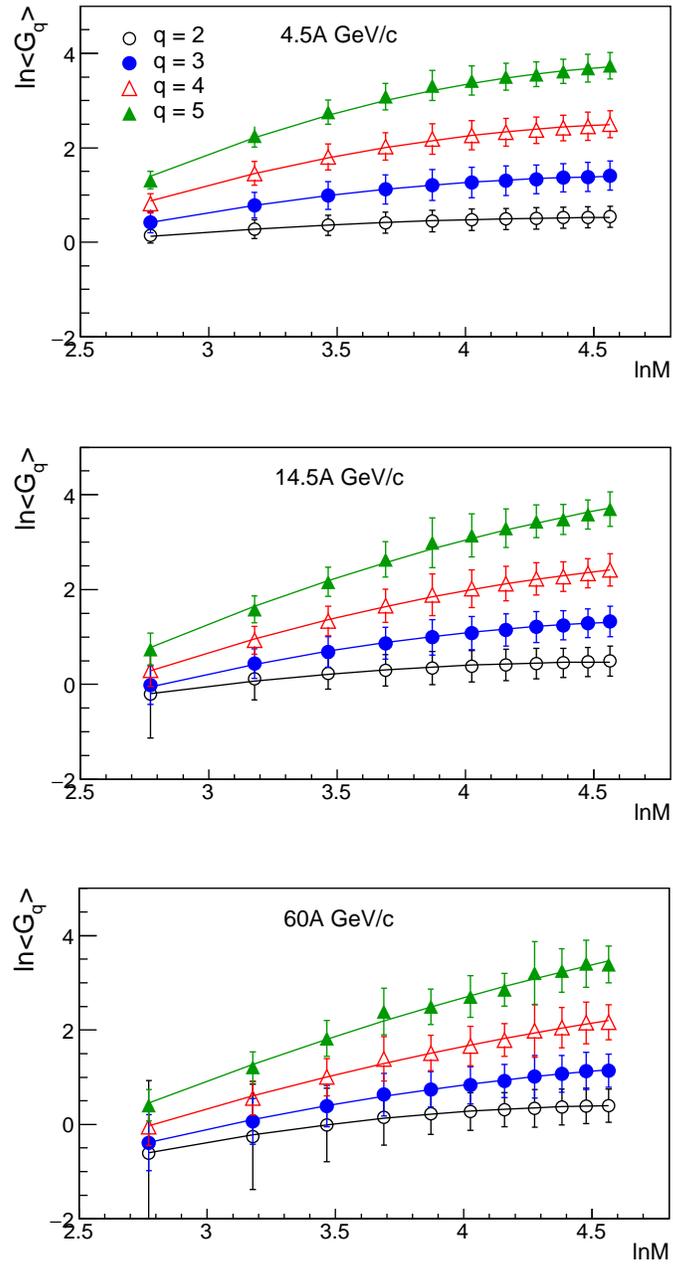,width=10cm}}
\vspace*{8pt}
\caption{Variations of \(ln<G_q>\) with \(lnM\) for the \am generated events at the three incident energies.}
\end{figure}
%
\newpage
\begin{figure}[th]
\centerline{\psfig{file=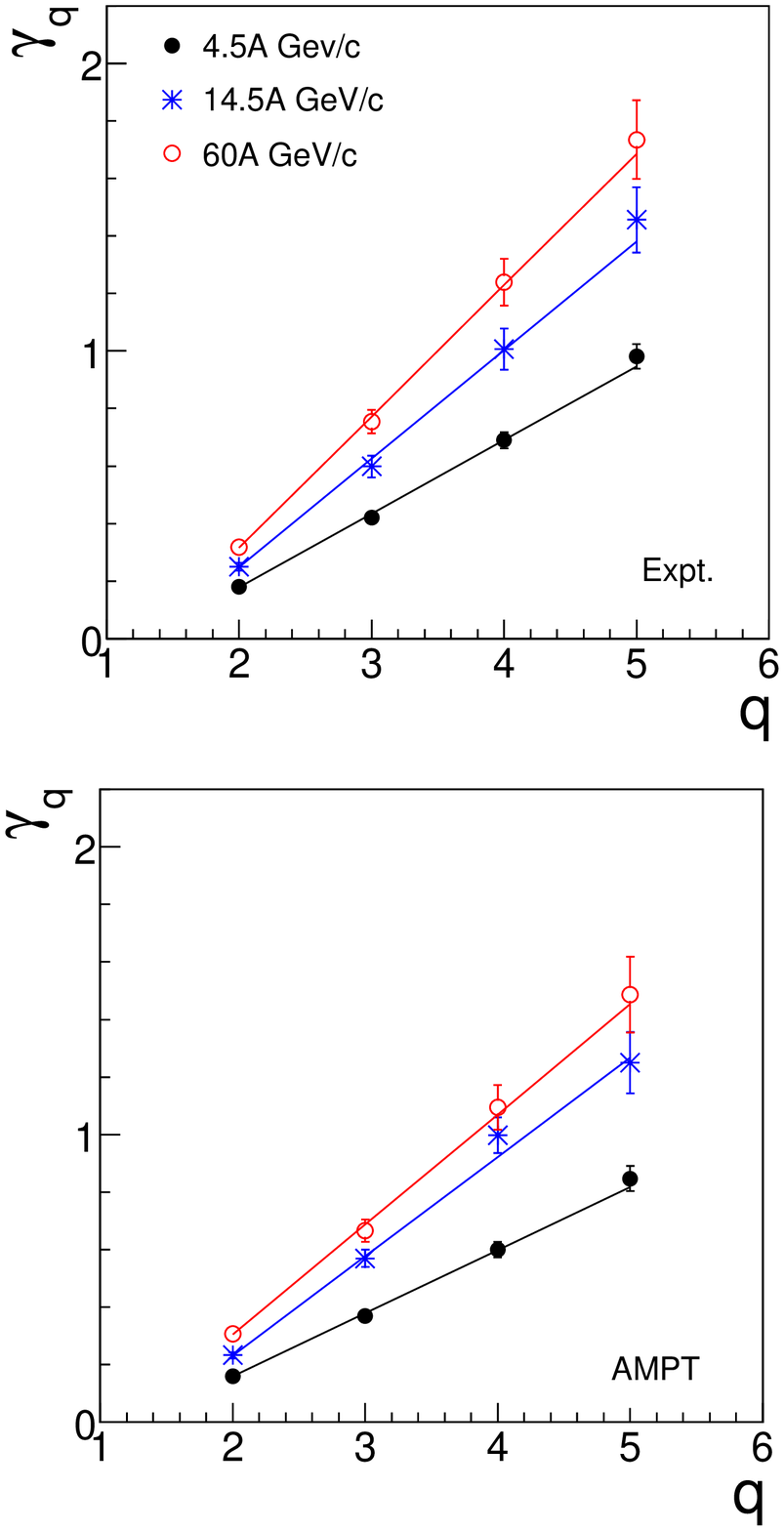,width=10cm}}
\vspace*{8pt}
\caption{Variations of \(\gamma_q\) with q for the experimental and \am simulated events at the three incident energies.}
\end{figure}
%
\newpage
\begin{figure}[th]
\centerline{\psfig{file=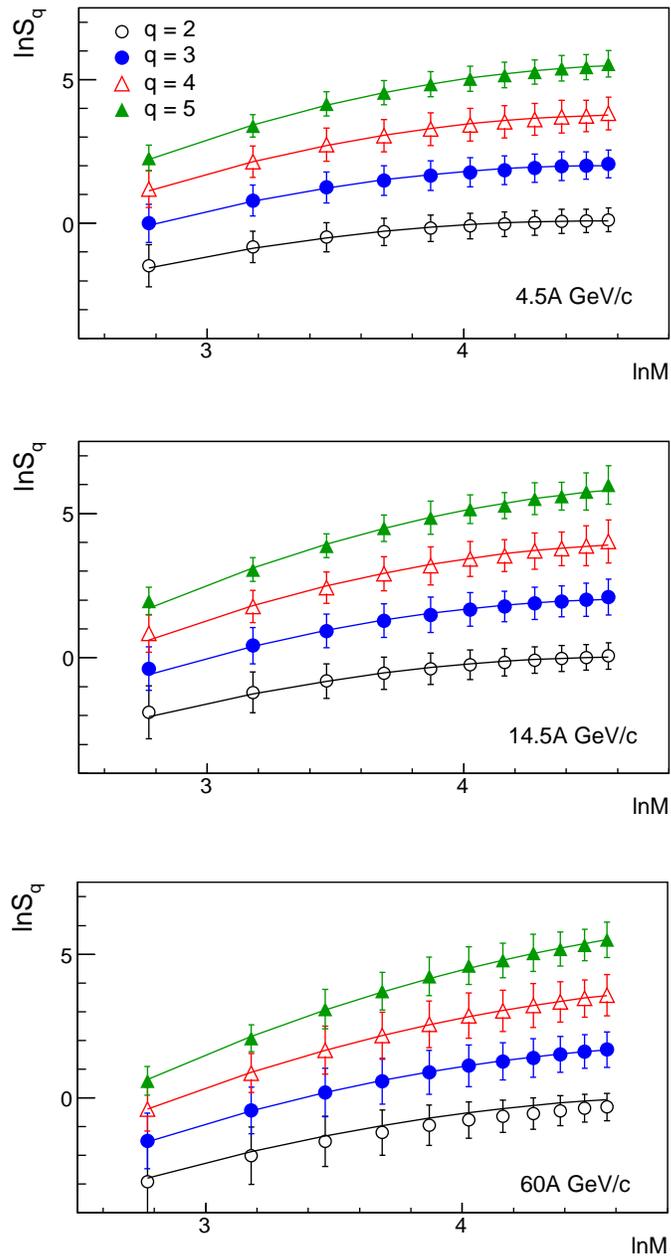,width=10cm}}
\vspace*{8pt}
\caption{Variations of \(lnS_q\) with \(lnM\) for the experimental 4.5, 14.5 and 60A GeV/c  $^{32}$S-AgBr interactions. }
\end{figure}
\newpage
\begin{figure}[th]
\centerline{\psfig{file=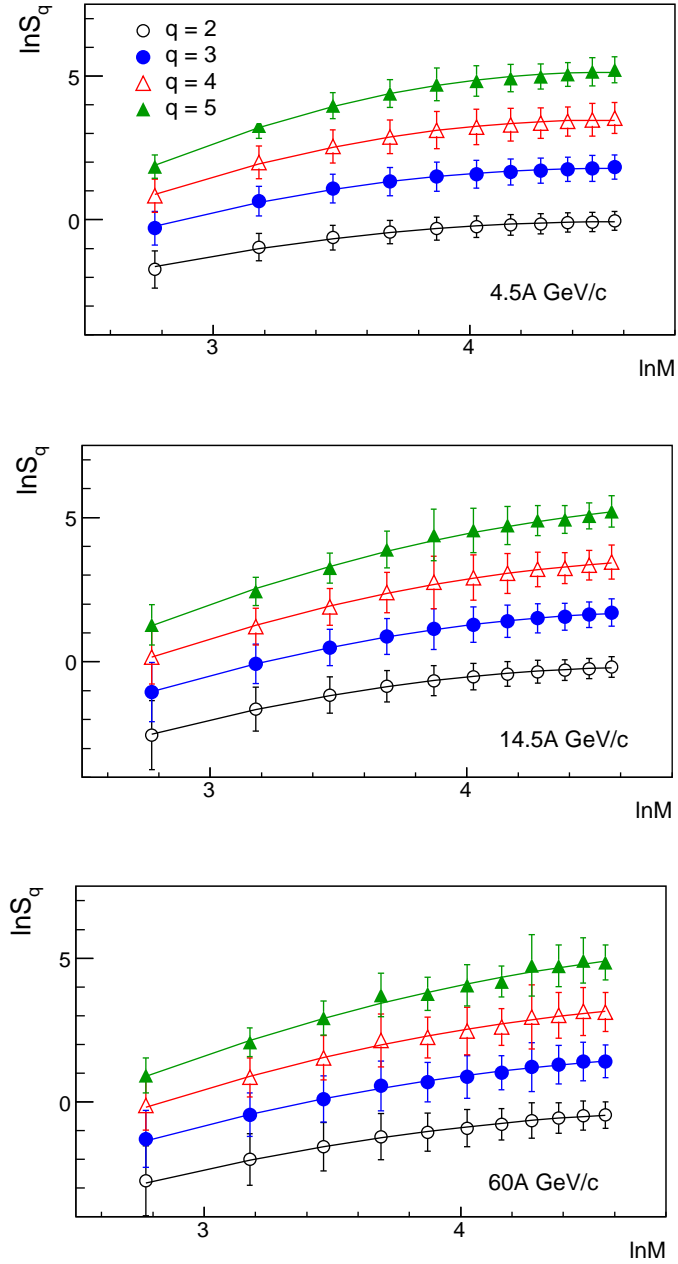,width=10cm}}
\vspace*{8pt}
\caption{Variations of \(lnS_q\) with \(lnM\) for the \am generated 4.5, 14.5 and 60A GeV/c  $^{32}$S-AgBr collisions.}
\end{figure}
%
\newpage
\begin{figure}[th]
\centerline{\psfig{file=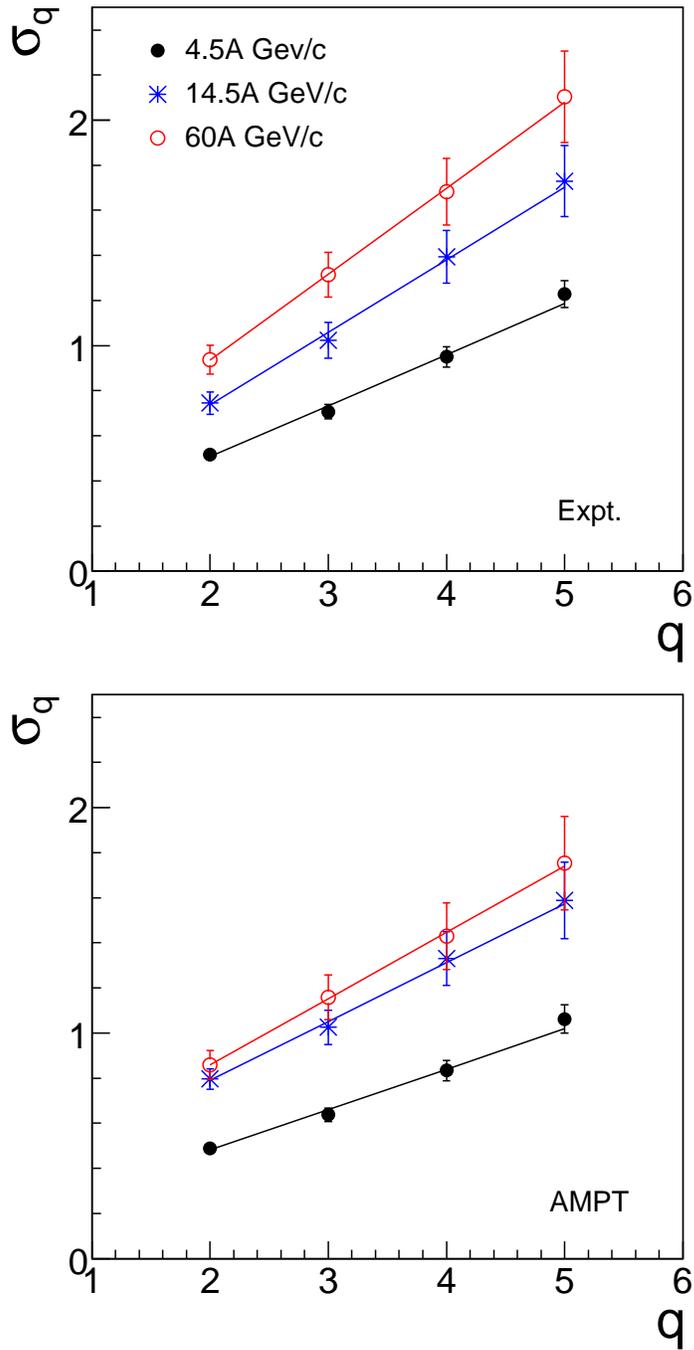,width=10cm}}
\vspace*{8pt}
\caption{Variations of \(\sigma_q\) with q for the experimental and \am events at the three incident energies.}
\end{figure}

\end{document}